\documentclass[10pt]{article}
\usepackage[colorlinks,
            linkcolor=green,
            urlcolor=blue,
            anchorcolor=red,
            citecolor=green
            ]{hyperref}
\usepackage{graphicx}
\usepackage{subfigure}
\usepackage{epstopdf}
\usepackage{rotating}
\usepackage{amsmath}
\usepackage{amssymb}
\usepackage{float}
\usepackage{color, soul}

\begin{document}
\begin{center}
{\Large \bf Chemical potentials of light hadrons and quarks from yield ratios of negative to positive particles in high energy $pp$ collisions}

\vskip1.0cm

Xing-Wei He$^{a}$, Hua-Rong Wei$^{b,}${\footnote{E-mail: huarongwei@qq.com;huarongwei@lsu.edu.cn}}, and Fu-Hu Liu$^{a,}${\footnote{E-mail: fuhuliu@163.com;fuhuliu@sxu.edu.cn}}

{\small\it $^a$Institute of Theoretical Physics \& State Key Laboratory of Quantum Optics and Quantum Optics Devices,\\ Shanxi University, Taiyuan, Shanxi 030006, China
$^b$Department of Optoelectric Engineering, Lishui University, Lishui, Zhejiang 323000, China}
\end{center}

\vskip1.0cm

{\bf Abstract:} We describe the transverse momentum spectra of $\pi^\pm$, $K^\pm$, $p$, and $\bar{p}$ produced in proton-proton ($pp$) collisions at different collision energies by using a Tsallis-Pareto-type function, and obtain the yield ratios, $k_{\pi}$, $k_{K}$, and $k_{p}$, of negative to positive particles from the fitting results of transverse momentum spectra and the
extracted normalization constants. The transverse momentum dependent and energy dependent chemical potentials of light hadrons ($\pi$, $K$, and $p$) and quarks ($u$, $d$, and $s$) in $pp$ collisions are then extracted successively from the yield ratios. The six types of transverse momentum dependent chemical potentials show the trend of being close to zero in low-$p_T$
region and away from zero in high-$p_T$ region. Meanwhile, the energy dependent chemical potentials seem to decrease slightly with the increase of the collision energy,
and the limiting values of the six types of chemical potentials are zero at very high energy in $pp$ collisions, which confirms that the partonic interactions possibly play an important role at RHIC and LHC, especially at LHC.
\\

\\


\vskip1.0cm

{\section{Introduction}}

The baryon- and quark-chemical potentials on the quantum
chromodynamics (QCD) phase diagram [1, 2] play an important role in
low-temperature and high-temperature systems. Lattice QCD (LQCD) [3--5], as a powerful tool to investigate the QCD matter in high-temperature system, indicates that the QCD transition is a crossover at small chemical potentials [6, 7], so the values of chemical potentials for high densities, though small, are not negligible. Some researches show that with the increase of collision energy over a range from GeV to TeV, the net baryon density at mid-rapidity decreases [8], the chemical freeze-out temperature of interacting system increases, and the limiting value of baryon chemical potential gradually becomes 1 MeV or zero [9--11]. Combining with other physical quantity, such as transition temperature, speed of sound, pressure and so on, one can obtain more information on the QCD matter and QCD phase diagram [7, 12--15]. It is therefore worthwhile to study the evolution and fluctuation of chemical potentials, and the dependencies of chemical potentials on other physical quantities. In particular, the study of the transverse momentum and center-of-mass energy dependent chemical potentials is useful in perfecting the QCD phase diagram.

The successfully running of the Relativistic Heavy Ion Collider (RHIC) in USA and the Large Hadron Collider (LHC) in Switzerland attracts more studies due to it's creating a high temperature and high density system, which goes through the phase transitions from hadronic matter to quark-gluon plasma (QGP) [16--18] and from QGP to hadronic matter [19, 20]. In gold-gold (Au-Au) collisions, when energy is above 39 GeV, the partonic interactions may play an important role, but when energy is below 11.5 GeV, the hadronic
interactions possibly play an important role [21--24]. The proton-proton ($pp$) collision system at high energy, a relatively simple collision system, is often used to be compared to complex collision systems like Au-Au, lead-lead (Pb-Pb) and so on. Due to the infamous sign problem in LQCD, calculations for non-vanishing chemical potentials are very difficult [25--28]. By using the yield ratios of negative to positive particles [29--31] to obtain chemical potential is an effective and simple research method. We are interested in studying the chemical potentials of different types of particles, and their dependencies on transverse momentum and center-of-mass energy in $pp$ collisions on the basis of yield ratios of negative to positive particles.

There are different ways to obtain the yield ratios of negative to positive particles. One can directly collect the experimental data of yield ratios measured by the productive international collaborations [10], which is a rapid and convenient method. One can also obtain the yield ratios from the consistent statistical law for transverse momentum spectra of different particles and extracted normalization constants [31]. Considering that the statistical law can describe the transverse momentum spectrum in a wider range, we think that one can obtain more accurate yield ratios and chemical potentials by using the method based on statistical law.

In this paper, we describe the transverse momentum ($p_{T}$) spectra of $\pi^\pm$, $K^\pm$, $p$, and $\bar{p}$ produced in $pp$ collisions over a center-of-mass energy ($\sqrt{s}$) range from 62.4 GeV to 13 TeV [32--34] using a Tsallis-Pareto-type function [34--36], and obtain the yield ratios, $k_{\pi}$, $k_{K}$, and $k_{p}$, of negative to positive particles. The $p_{T}$-dependent and $\sqrt{s}$-dependent chemical potentials of light hadrons ($\pi$, $K$, and $p$) and quarks ($u$, $d$, and $s$) in $pp$ collisions are then extracted from the yield ratios.

{\section{The model and formulism}}

The chemical potentials of some light hadrons and quarks are extracted from the yield ratios of negative to positive particles. The yield ratios are obtained from the normalization constants in describing the transverse momentum spectra of positive and negative particles with the Tsallis-Pareto-type function.

The Tsallis-Pareto-type function [34--36], which empirically describes both the low-$p_{T}$ exponential and the high-$p_{T}$ power-law behaviors, has a good reproduction of many measurements of particle spectra [33, 37, 38], and has the following form of
\begin{equation}
\frac{\mathrm{d}^{2}N}{\mathrm{d}y\mathrm{d}p_T}=\frac{\mathrm{d}N}{\mathrm{d}y}Cp_T
\bigg[1+\frac{m_{T}-m_{0}}{nT}\bigg]^{-n},
\end{equation}
where
\begin{equation}
C=\frac{(n-1)(n-2)}{nT[nT+(n-2)m_{0}]},
\end{equation}
$m_{T}=\sqrt{m_{0}^{2}+p_T^{2}}$, $m_0$ is the rest mass, $y$ is the rapidity, and $N$ is the number of particles. According to some non-extensive thermodynamics models of particle production [36], the free parameter $T$, which is connected with the average particle energy, represents the mean effective temperature of interacting system, $dN/dy$ characterizes the integrated yield, and $n$ reveals the ¡°non-extensivity¡± of the process, i.e. the departure of the spectra from the Boltzmann distribution.

Because of the lifetimes of particles contained top quark being too short to measure, and the limited data being collected, only the chemical potentials of some light hadrons such as $\pi$, $K$, and $p$, as well as some light quarks such as $u$, $d$, and $s$ are obtained in the present work. Within the thermal and statistical model [29], according to the statistical arguments
based on the chemical and thermal equilibrium, the relations between antiparticle to particle (negative to positive particle) yield ratios and chemical potentials of hadrons can be written as [29--31]
\begin{gather}
k_{\pi}=\exp\bigg(-\frac{2\mu_{\pi}}{T_{ch}}\bigg),\notag\\
k_{K}=\exp\bigg(-\frac{2\mu_{K}}{T_{ch}}\bigg),\notag\\
k_{p}=\exp\bigg(-\frac{2\mu_{p}}{T_{ch}}\bigg),
\end{gather}
where $k_{\pi}$, $k_{K}$, and $k_{p}$ denote the yield ratios of antiparticles, $\pi^{-}$, $K^{-}$, and $\overline{p}$, to particles, $\pi^{+}$, $K^{+}$, and $p$, respectively. And
$\mu_{\pi}$, $\mu_{K}$, and $\mu_{p}$ represent the chemical potentials of $\pi$, $K$, and $p$, respectively. In the framework of a statistical thermal model of non-interacting gas particles with the assumption of standard Maxwell-Boltzmann statistics [16,17, 39], the chemical freeze-out temperature of the interacting system can be obtained and written as
\begin{equation}
T_{ch}=T_{\lim}\frac{1}{1+\exp[2.60-\ln(\sqrt{s_{NN}})/0.45]},
\end{equation}
where the ``limiting" temperature $T_{\lim}$ is 0.164 GeV, and $\sqrt{s_{NN}}$ is in the unit of GeV [14, 39].

Based on the above chemical freeze-out temperature and refs. [10, 31, 40], one can obtain the yield ratios in terms of quark chemical potentials to be
\begin{gather}
k_{\pi}=\exp\bigg[-\frac{(\mu_{u}-\mu_{d})}{T_{ch}}\bigg]\bigg/\exp\bigg[\frac{(\mu_{u}-\mu_{d})}{T_{ch}}\bigg]=\exp\bigg[-\frac{2(\mu_{u}-\mu_{d})}{T_{ch}}\bigg],\notag\\
k_{K}=\exp\bigg[-\frac{(\mu_{u}-\mu_{s})}{T_{ch}}\bigg]\bigg/\exp\bigg[\frac{(\mu_{u}-\mu_{s})}{T_{ch}}\bigg]=\exp\bigg[-\frac{2(\mu_{u}-\mu_{s})}{T_{ch}}\bigg],\notag\\
k_{p}=\exp\bigg[-\frac{(2\mu_{u}+\mu_{d})}{T_{ch}}\bigg]\bigg/\exp\bigg[\frac{(2\mu_{u}+\mu_{d})}{T_{ch}}\bigg]=\exp\bigg[-\frac{2(2\mu_{u}+\mu_{d})}{T_{ch}}\bigg],
\end{gather}
where $\mu_{u}$, $\mu_{d}$, and $\mu_{s}$ represent the chemical potentials of $u$, $d$, and $s$ quarks, respectively.

According to Eqs. (3) and (5), the chemical potentials of hadrons and quarks in terms of yield ratios can be obtained respectively, and are as follows:
\begin{gather}
\mu_{\pi}=-\frac{1}{2}T_{ch}\cdot\ln(k_{\pi}),\notag\\
\mu_{K}=-\frac{1}{2}T_{ch}\cdot\ln(k_{K}),\notag\\
\mu_{p}=-\frac{1}{2}T_{ch}\cdot\ln(k_{p}),
\end{gather}
and
\begin{gather}
\mu_{u}=-\frac{1}{6}T_{ch}\cdot\ln(k_{\pi}\cdot k_{p}),\notag\\
\mu_{d}=-\frac{1}{6}T_{ch}\cdot\ln(k_{K}^{-2}\cdot k_{p}),\notag\\
\mu_{s}=-\frac{1}{6}T_{ch}\cdot\ln(k_{\pi}\cdot k_{K}^{-3}\cdot
k_{p}).
\end{gather}

By describing the $p_{T}$ spectra of $\pi^\pm$, $K^\pm$, $p$, and $\bar{p}$ in $\sqrt{s}=$ 62.4, 200, 900, 2760, 7000 and 13000 GeV $pp$ collisions using the Tsallis-Pareto-type function, we can obtain the chemical potentials of light hadrons ($\pi$, $K$, and $p$) and light quarks ($u$, $d$, and $s$). Then, the dependencies of chemical potentials on $p_{T}$ and $\sqrt{s}$ can be analyzed.

It should be noted that the above discussions are based on the assumption of thermal and statistical equilibrium and chemical equilibrium. Although $pp$ collisions are very small, the multiplicity of particles produced in the collisions at high energy is enough high. Heavy ion collisions at high energy result in higher multiplicity. These mean that the assumption of thermal and statistical equilibrium and chemical equilibrium is suitable. The above formulas are valid simultaneously in the environments of high energy $pp$ and heavy-ion collisions.

{\section{Results and discussion}}

Figure 1 presents the transverse momentum spectra of $\pi^\pm$, $K^\pm$, $p$, and $\bar{p}$ produced in $pp$ collisions at $\sqrt{s}=$ (a)(d) 62.4, (b)(e) 200, and (c)(f) 900 GeV, where
$E$, $\sigma$, and $N_{ev}$ on the vertical axis denote the particle energy, cross-section, and corrected number of inelastic $pp$ collisions in the data sample, respectively. The data
measured by (a)(b)(d)(e) the PHENIX Collaboration in the mid-pseudorapidity range $|\eta|<0.35$ [32], and (c)(f) the CMS Collaboration in the mid-rapidity range $|y|<1$ [33], are represented in different panels by different symbols. The error bars, are statistical only for 62.4 and 200 GeV $pp$ collisions, and are the combined statistical and systematic ones of 3.0\% for 900 GeV $pp$ collisions. The curves are our results fitted by using the Tsallis-Pareto-type function. The values of normalization constant ($N_0$) for comparison between the curve
and data, free parameters ($T$ and $n$) for the fit, $\chi^2$, and number of degree of freedom (dof) are given in Table 1. One can see that all the fitting results by using the Tsallis-Pareto-type function are consistent with the experimental data. $T$ increases with increase of particle mass and collision energy, and $N_0$ decreases with increase of particle mass. It should be noted that the particle yield ratio is represented by $N_0$ of the spectra of negative and positive particles. The relative value of $N_0$ is enough to obtain the particle yield ratio.

\begin{figure}[H]
\hskip-0.0cm {\centering
\includegraphics[width=16.0cm]{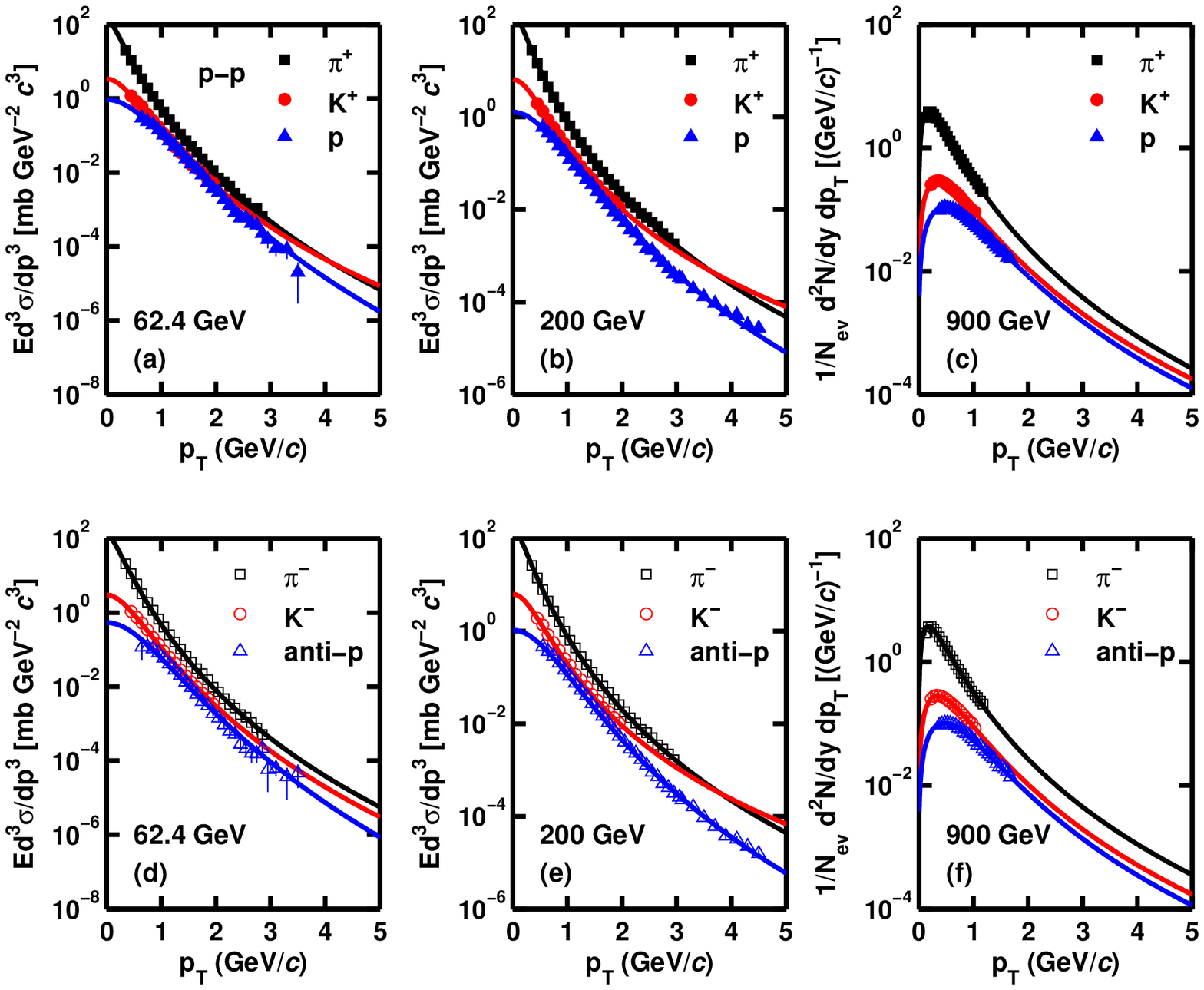}}
\vskip-0.18cm Figure 1. Transverse momentum spectra for (a)(b)(c) positive ($\pi^+$, $K^+$, $p$) and (d)(e)(f) negative ($\pi^-$, $K^-$, $\bar{p}$) particles produced in $pp$ collisions at
$\sqrt{s}=$ (a)(d) 62.4, (b)(e) 200, and (c)(f) 900 GeV. The experimental data represented by the symbols are measured by the PHENIX Collaboration in $|\eta|<0.35$ at 62.4 and 200 GeV [32],
and by the CMS Collaboration in $|y|<1$ at 900 GeV [33]. The errors, are statistical only in Figs. 1(a), 1(b), 1(d), and 1(e) , and include statistical and systematic errors (3.0\%) in Figs.1(c) and 1(f). The curves are fits by the Tsallis-Pareto-type function.
\end{figure}

Figure 2 shows the same as Fig. 1, but for $\sqrt{s}=$ (a)(d) 2.76, (b)(e) 7, and (c)(f) 13 TeV. The symbols also denote the experimental data recorded by the CMS collaboration in the range
$|y|<1$ [33, 34]. The error bars indicate the combined uncorrelated statistical and systematic uncertainties, and the fully correlated normalization uncertainty is 3.0\%. The curves
are our results fitted by using the Tsallis-Pareto-type function. The values of $N_0$, $T$, $n$, $\chi^2$, and dof are displayed in Tables 1. It is not hard to see that the experimental data can be well fitted by the Tsallis-Pareto-type function. Similarly, $T$ increases with increase of particle mass and collision energy, and $N_0$ decreases with increase of particle mass.

Generally, the above fitting results are approximately independent of models. In fact, only the normalization constants are used in the extractions of chemical potentials. One can even use the normalization constants from the data directly [10]. In the present work, we use the fitting results instead of the data due to the fact that the data in high $p_T$ region in some cases are not available. In the case of using other models such as the Tsallis distribution [41,42], the results are the same, if not equal to each other. In fact, in terms of entropy index $q$ in the Tsallis distribution, we have $n=1/(q-1)$ in Eq. (1) when $p_T\gg m_0$. This means that the Tsallis-Pareto-type function [34--36] is equivalent to the Tsallis distribution in most cases.

We use the Tsallis-Pareto-type function and give up to use the Tsallis distribution in the present work due to the fact that the former has few applications and the latter has many applications in the community. In particular, we have used the Tsallis distribution in our previous works [43--45]. To show the variousness in the fitting functions, we use the Tsallis-Pareto-type function in the present work. In like manner, one can see from Table 1 that the temperature parameter increases usually with the collision energy, and the entropy index does not change obviously with the collision energy within the statistical errors.

According to the fitting results and extracted normalization constants from the above comparisons, the yield ratios of negative to positive particles, $k_{\pi}$, $k_{K}$, and $k_{p}$ versus
transverse momentum and collision energy are obtained though the yield is treated as the normalization constant which is almost independent of models. In other words, the model can explain altogether yields and particle ratios measured in experiments. On the basis of the derivative yield ratios and Eqs. (6) and (7), the hadron chemical potentials, $\mu_{\pi}$, $\mu_{K}$, and $\mu_{p}$ of $\pi$, $K$, and $p$, and quark chemical potentials, $\mu_{u}$, $\mu_{d}$, and $\mu_{s}$ of $u$, $d$, and $s$ quarks, which vary with transverse momentum and collision energy, are obtained and shown in Figs. 3--5, respectively. We shall describe Figs. 3--5 one by one.

\begin{figure}[H]
\hskip-0.0cm {\centering
\includegraphics[width=16.0cm]{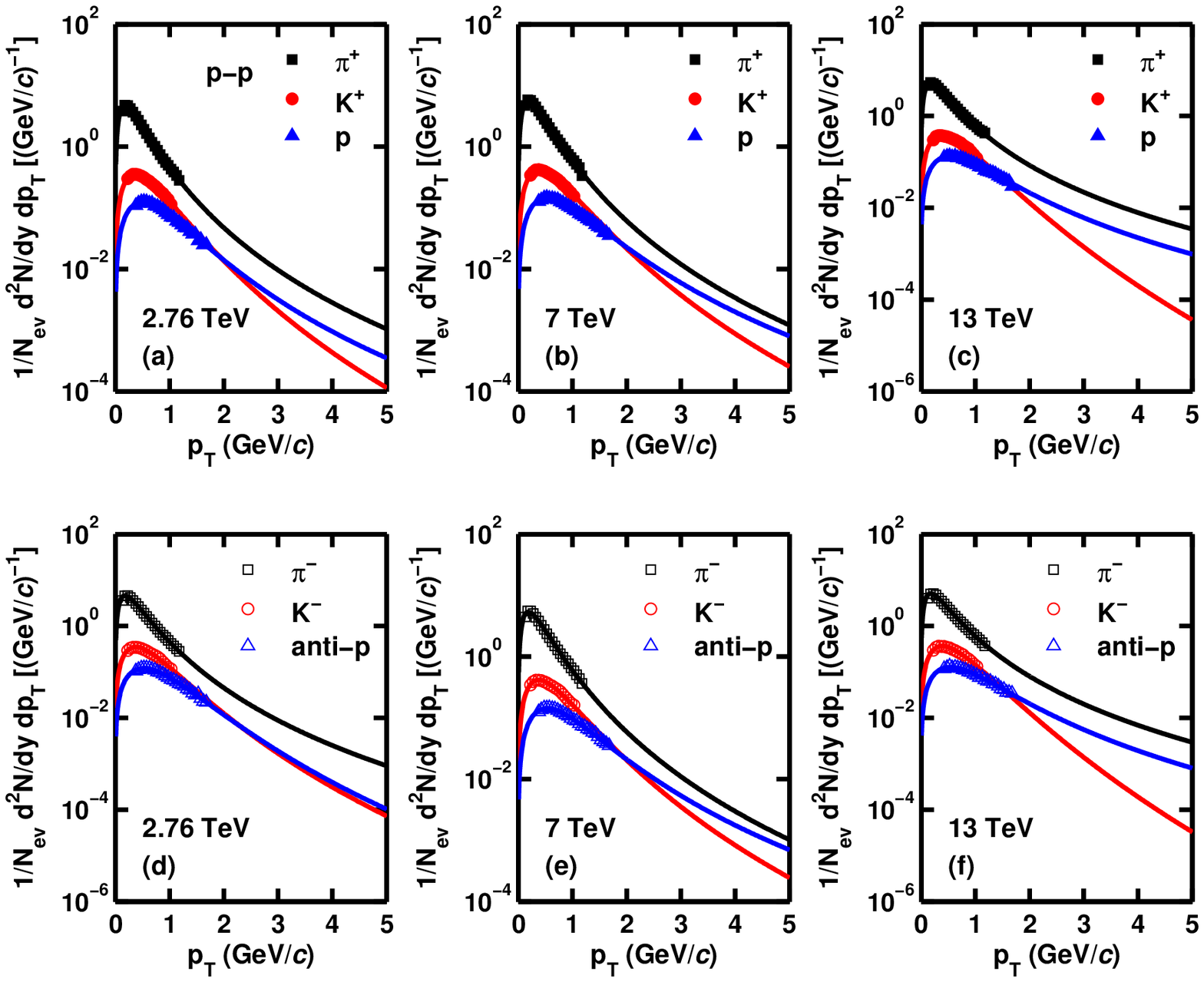}}
\vskip-0.18cm  Figure 2. Transverse momentum spectra for (a)(b)(c) positive ($\pi^+$, $K^+$, $p$) and (d)(e)(f) negative ($\pi^-$, $K^-$, $\bar{p}$) particles produced in $pp$ collisions at
$\sqrt{s}=$ (a)(d) 2.76, (b)(e) 7, and (c)(f) 13 TeV. The symbols represent the experimental data recorded by the CMS Collaboration in $|y|<1$ [33, 34]. The errors are the combined uncorrelated statistical and systematic ones, and the fully correlated normalization uncertainty is 3.0\%. The curves are our results fitted by the Tsallis-Pareto-type function.
\end{figure}

Figure 3 displays the transverse momentum dependent $\mu_{\pi}$, $\mu_{K}$, and $\mu_{p}$ of $\pi$, $K$, and $p$ produced in $pp$ collisions at different energies shown in panels (a)--(f). The curves are our results calculated by using Eq. (6). One can see that, at the same energy, the three curves corresponding to three types of chemical potentials show different behaviors with increase of transverse momentum, and at different energies, the curves corresponding to any type of chemical potentials also show different behaviors. Though the curves fluctuate obviously, the values of $\mu_{\pi}$, $\mu_{K}$, and $\mu_{p}$ are not too large, and most of them are within $\pm 50$ MeV. Moreover, it is not hard to find that the three curves concentrate in low-$p_T$ region and diverge in high-$p_T$ region. In particular, when the collision energy is greater than or equal to 900 GeV, the three curves are obviously close to zero in low-$p_T$ region, and when the energy is 7 TeV, $\mu_{\pi}$, $\mu_{K}$, and $\mu_{p}$ focus on around zero in the whole transverse momentum region.

Figure 4 is the same as Fig. 3, but for the quark chemical potentials, $\mu_{u}$, $\mu_{d}$, and $\mu_{s}$ of $u$, $d$, and $s$. The curves are our results calculated by using Eq. (7). we
can see that, at the same energy in the range from 62.4 GeV to 13 TeV, the curves of $\mu_{u}$, $\mu_{d}$, and $\mu_{s}$ show different trends with increase of transverse momentum. At different energies, the three curves of quark chemical potentials also show different trends. Similarly, the most values of $\mu_{u}$, $\mu_{d}$, and $\mu_{s}$ are within $\pm 50$ MeV. And
the three curves of quark chemical potentials concentrate in low-$p_T$ region and diverge in high-$p_T$ region. Particularly, when the collision energy is greater than or equal to 900 GeV, the values of $\mu_{u}$, $\mu_{d}$, and $\mu_{s}$ in low-$p_T$ region are nearly equal to zero, and when the energy is $7$ TeV, all the values of $\mu_{u}$, $\mu_{d}$, and $\mu_{s}$ focus on around zero in the whole transverse momentum region. There are some similarities between the $p_T$-dependent chemical potentials of light hadrons and quarks in $pp$ collisions over an energy range from the RHIC to LHC.

\vspace{0.75cm}

{\scriptsize {Table 1. Values of normalization constant, free parameters, $\chi^2$, and dof corresponding to Tsallis-Pareto-type function in Figs. 1 and 2.
{%
\begin{center}
\begin{tabular}{cccccccc}
\hline \hline
Figure  &Type   &Particle   &$N_0$    &$T$ (GeV)    &$n$    &$\chi^2$    &dof\\
\hline
   Figure 1(a) &62.4 GeV&$\pi^{+}$     &1.042 $\pm$ 0.042 &0.122 $\pm$ 0.003 &      10.9 $\pm$ 0.4&1.079&23\\
               &        &$K^{+}$       &0.089 $\pm$ 0.004 &0.160 $\pm$ 0.005 & \,\,\,9.1 $\pm$ 0.6&0.577&13\\
               &        &$p$           &0.044 $\pm$ 0.004 &0.180 $\pm$ 0.007 &      12.8 $\pm$ 1.0&4.819&24\\
   Figure 1(d) &        &$\pi^{-}$     &0.995 $\pm$ 0.004 &0.120 $\pm$ 0.002 &      10.9 $\pm$ 0.4&2.107&23\\
               &        &$K^{-}$       &0.078 $\pm$ 0.004 &0.162 $\pm$ 0.004 &      10.8 $\pm$ 0.5&0.285&13\\
               &        &$\overline{p}$&0.025 $\pm$ 0.002 &0.180 $\pm$ 0.008 &      13.2 $\pm$ 1.2&6.291&24\\
\hline
   Figure 1(b) &200 GeV &$\pi^{+}$     &1.176 $\pm$ 0.068 &0.117 $\pm$ 0.003 & \,\,\,8.5 $\pm$ 0.2&0.695&24\\
               &        &$K^{+}$       &0.131 $\pm$ 0.004 &0.135 $\pm$ 0.004 & \,\,\,6.0 $\pm$ 0.2&0.187&13\\
               &        &$p$           &0.052 $\pm$ 0.005 &0.176 $\pm$ 0.006 & \,\,\,9.7 $\pm$ 0.5&2.407&31\\
   Figure 1(e) &        &$\pi^{-}$     &1.131 $\pm$ 0.070 &0.117 $\pm$ 0.002 & \,\,\,8.6 $\pm$ 0.2&0.852&24\\
               &        &$K^{-}$       &0.124 $\pm$ 0.004 &0.132 $\pm$ 0.003 & \,\,\,6.0 $\pm$ 0.2&0.974&13\\
               &        &$\overline{p}$&0.042 $\pm$ 0.004 &0.174 $\pm$ 0.004 & \,\,\,9.8 $\pm$ 0.6&1.632&31\\
\hline
   Figure 1(c) &900 GeV &$\pi^{+}$     &3.826 $\pm$ 0.094 &0.127 $\pm$ 0.004 & \,\,\,7.5 $\pm$ 0.3&0.505&19\\
               &        &$K^{+}$       &0.476 $\pm$ 0.020 &0.184 $\pm$ 0.009 & \,\,\,7.1 $\pm$ 0.6&0.083&14\\
               &        &$p$           &0.222 $\pm$ 0.008 &0.197 $\pm$ 0.011 & \,\,\,6.9 $\pm$ 1.0&0.369&24\\
   Figure 1(f) &        &$\pi^{-}$     &3.770 $\pm$ 0.130 &0.125 $\pm$ 0.004 & \,\,\,7.1 $\pm$ 0.7&0.927&19\\
               &        &$K^{-}$       &0.464 $\pm$ 0.016 &0.180 $\pm$ 0.008 & \,\,\,7.0 $\pm$ 0.5&0.151&14\\
               &        &$\overline{p}$&0.206 $\pm$ 0.008 &0.191 $\pm$ 0.012 & \,\,\,6.8 $\pm$ 1.0&0.837&24\\
\hline
   Figure 2(a) &2.76 TeV&$\pi^{+}$     &4.852 $\pm$ 0.228 &0.122 $\pm$ 0.006 & \,\,\,6.2 $\pm$ 0.4&0.511&19\\
               &        &$K^{+}$       &0.602 $\pm$ 0.018 &0.207 $\pm$ 0.009 & \,\,\,9.2 $\pm$ 1.0&0.104&14\\
               &        &$p$           &0.280 $\pm$ 0.010 &0.218 $\pm$ 0.008 & \,\,\,6.2 $\pm$ 0.8&0.989&24\\
   Figure 2(d) &        &$\pi^{-}$     &4.734 $\pm$ 0.214 &0.125 $\pm$ 0.004 & \,\,\,6.4 $\pm$ 0.3&0.641&19\\
               &        &$K^{-}$       &0.586 $\pm$ 0.022 &0.213 $\pm$ 0.008 &      10.5 $\pm$ 1.7&0.236&14\\
               &        &$\overline{p}$&0.262 $\pm$ 0.012 &0.239 $\pm$ 0.011 & \,\,\,9.3 $\pm$ 0.7&1.056&24\\
\hline
   Figure 2(b) &7 TeV   &$\pi^{+}$     &5.914 $\pm$ 0.284 &0.131 $\pm$ 0.007 & \,\,\,6.5 $\pm$ 0.6&0.976&19\\
               &        &$K^{+}$       &0.750 $\pm$ 0.022 &0.201 $\pm$ 0.006 & \,\,\,8.7 $\pm$ 0.7&0.066&14\\
               &        &$p$           &0.364 $\pm$ 0.012 &0.243 $\pm$ 0.008 & \,\,\,5.9 $\pm$ 0.4&0.426&24\\
   Figure 2(e) &        &$\pi^{-}$     &5.810 $\pm$ 0.212 &0.135 $\pm$ 0.008 & \,\,\,6.8 $\pm$ 0.6&0.962&19\\
               &        &$K^{-}$       &0.740 $\pm$ 0.020 &0.216 $\pm$ 0.007 & \,\,\,7.4 $\pm$ 0.8&0.138&14\\
               &        &$\overline{p}$&0.348 $\pm$ 0.014 &0.239 $\pm$ 0.014 & \,\,\,5.9 $\pm$ 0.6&0.519&24\\
\hline
   Figure 2(c) &13 TeV  &$\pi^{+}$     &5.664 $\pm$ 0.224 &0.119 $\pm$ 0.006 & \,\,\,5.2 $\pm$ 0.4&0.294&19\\
               &        &$K^{+}$       &0.638 $\pm$ 0.020 &0.229 $\pm$ 0.014 &      14.1 $\pm$ 3.2&0.167&14\\
               &        &$p$           &0.334 $\pm$ 0.014 &0.234 $\pm$ 0.011 & \,\,\,5.2 $\pm$ 0.5&0.540&23\\
   Figure 2(f) &        &$\pi^{-}$     &5.600 $\pm$ 0.200 &0.123 $\pm$ 0.005 & \,\,\,5.4 $\pm$ 0.5&1.106&19\\
               &        &$K^{-}$       &0.630 $\pm$ 0.018 &0.234 $\pm$ 0.014 &      15.0 $\pm$ 3.0&0.099&14\\
               &        &$\overline{p}$&0.320 $\pm$ 0.010 &0.242 $\pm$ 0.017 & \,\,\,5.6 $\pm$ 0.7&0.425&23\\
\hline
\end{tabular}%
\end{center}
}} }

\vspace{0.75cm}

The $\sqrt{s}$-dependent chemical potentials of (a) light hadrons of $\pi$, $K$, and $p$, and (b) quarks of $u$, $d$, and $s$, which are independent of transverse momentum, are exhibited in Fig. 5. The symbols are the calculated results according to normalization constants and Eqs. (6) and (7). To see clearly the dependencies of different types of $\mu$ on $\sqrt{s}$, we fit the calculated results with two kinds of functions. The solid curves are the fits according to the below empirical functions of
\begin{gather}
\mu_{\pi}=(5\pm2)(\sqrt{s})^{-(0.162\pm0.070)},\notag\\
\mu_{K}=(27\pm1)(\sqrt{s})^{-(0.352\pm0.054)},\notag\\
\mu_{p}=(198\pm2)(\sqrt{s})^{-(0.454\pm0.059)},\notag\\
\mu_{u}=(61\pm2)(\sqrt{s})^{-(0.412\pm0.056)},\notag\\
\mu_{d}=(131\pm2)(\sqrt{s})^{-(0.668\pm0.116)},\notag\\
\mu_{s}=(37\pm4)(\sqrt{s})^{-(0.538\pm0.178)},
\end{gather}
with $\chi^2$/dof to be 0.154/3, 0.246/3, 0.643/3, 0.523/3, 0.550/3, and 0.457/3, respectively, where $\mu$ and $\sqrt{s}$ are in the units of MeV and GeV respectively. And the dashed curves are the fits according to the below empirical functions of
\begin{gather}
\mu_{\pi}=(12\pm2)(\ln{\sqrt{s}})^{-(1.017\pm0.471)},\notag\\
\mu_{K}=(188\pm2)(\ln{\sqrt{s}})^{-(2.305\pm0.307)},\notag\\
\mu_{p}=(2515\pm2)(\ln{\sqrt{s}})^{-(3.005\pm0.227)},\notag\\
\mu_{u}=(602\pm4)(\ln{\sqrt{s}})^{-(2.719\pm0.241)},\notag\\
\mu_{d}=(5539\pm3)(\ln{\sqrt{s}})^{-(4.419\pm0.600)},\notag\\
\mu_{s}=(901\pm7)(\ln{\sqrt{s}})^{-(3.649\pm1.013)},
\end{gather}
with $\chi^2$/dof to be 0.166/3, 0.148/3, 0.190/3, 0.186/3, 0.162/3, and 0.299/3, respectively. One can see that, the six types of chemical potentials seem to decrease slightly with increase of $\sqrt{s}$, and the limiting values of the six types of chemical potentials are zero at very high energy in $pp$ collisions. Meanwhile, the fluctuations in all the curves corresponding to $\sqrt{s}$-dependent chemical potentials are smaller than those corresponding to $p_T$-dependent chemical potentials, which means that the dependence of chemical potential on collision energy is stronger than that on transverse momentum. Interestingly, the value of any type of $\sqrt{s}$-dependent chemical potential at any energy, is less than the mean value of chemical potentials in the whole transverse momentum region of the corresponding particle at the corresponding energy.

Equations (8) and (9) can be translated into the following forms of
\begin{gather}
\ln{\mu}=a+b\ln{\sqrt{s}},
\end{gather}
and
\begin{gather}
\ln{\mu}=c+d\ln{(\ln{\sqrt{s}})},
\end{gather}
respectively, where parameters $a$, $b$, $c$, and $d$ can be obtained from Eqs. (8) and (9). In Eq. (10), $\ln{\mu}$ varies linearly with  $\ln{\sqrt{s}}$, while in Eq. (11), $\ln{\mu}$ varies linearly with  $\ln{(\ln{\sqrt{s}})}$. In low energy range, $\ln{\mu}$ decreases with increase of $\ln{\sqrt{s}}$ in Eq. (11) more rapidly than that in Eq. (10), which renders that the differences between some calculated values ($\mu_{p}$, $\mu_{d}$, and $\mu_{s}$) obtained from Eq. (10) at 62.4 GeV and corresponding data are relatively large. While at high energy range, some curves ($\mu_{d}$, and $\mu_{s}$) from Eq. (10) are more close to corresponding data than those from Eq. (11). In conclusion, from the description results, both of the two fittings are generally acceptable in the energy range involved.

\begin{figure}[H]
\hskip-0.0cm {\centering
\includegraphics[width=16.0cm]{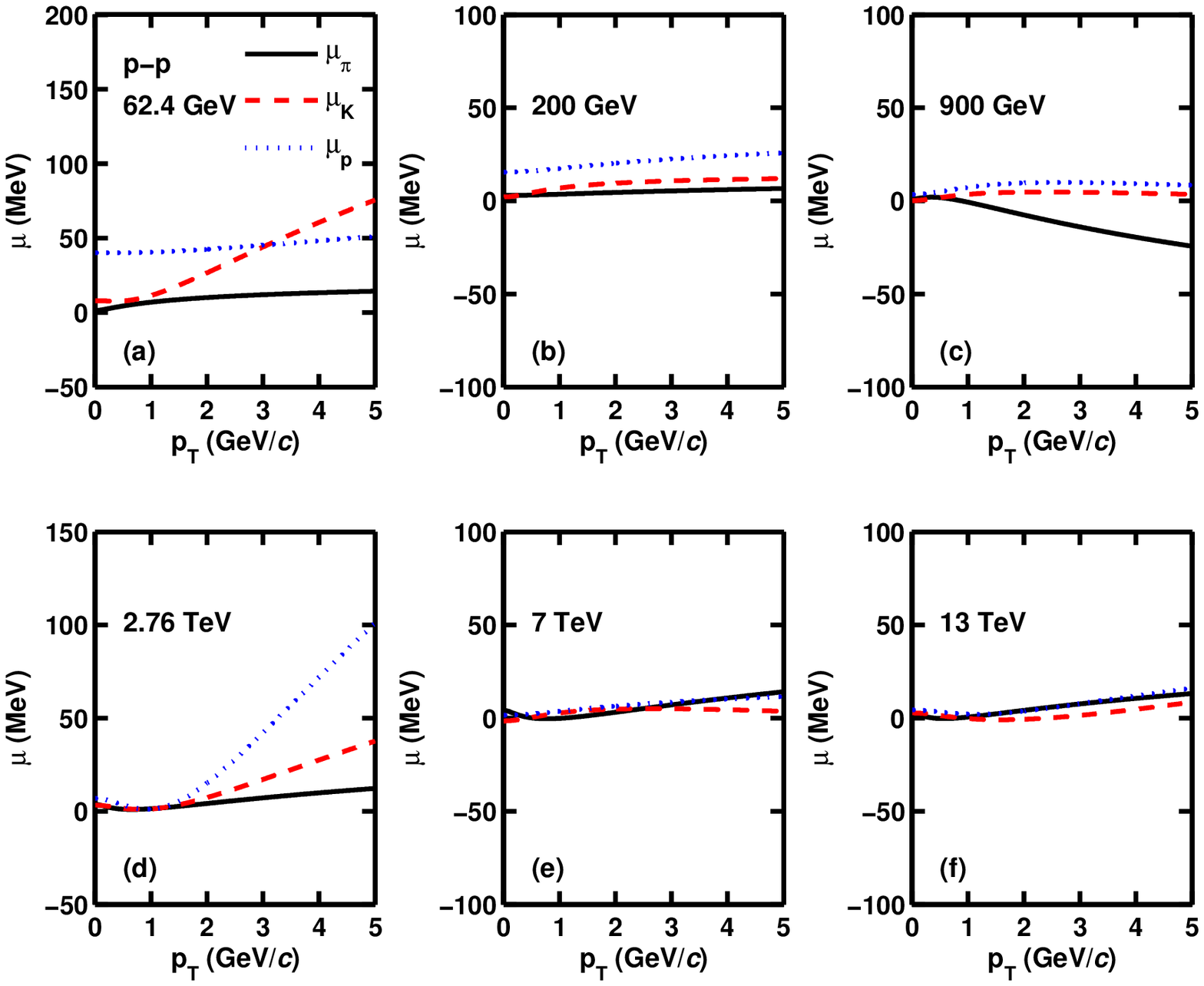}}
\vskip-0.18cm  Figure 3. Transverse momentum dependent chemical potentials, $\mu_{\pi}$, $\mu_{K}$, and $\mu_{p}$ of $\pi$, $K$, and $p$ produced in $pp$ collisions at different energies shown in panels (a)--(f). The curves are our results calculated by using Eq. (6).
\end{figure}

In Figs. 3 and 4, from most values of $\mu_{\pi}$, $\mu_{K}$, $\mu_{p}$, $\mu_{u}$, $\mu_{d}$, and $\mu_{s}$ being within $\pm50$ MeV, we can know that most values of the six yield ratios
(three negative to positive particle yield ratios, and three negative to positive quark yield ratios) are between 0.74 and 1.36. The six types of chemical potentials show different behaviors with increase of transverse momentum because of the different energies and different particle types, which is related to the yield ratios in different $p_T$ regions and results in so complex behaviors that they cannot be described in terms of uniform rule. Of course, we have to point out that, these curves showing large fluctuations with $p_T$, especially in extremely low- and high-$p_T$ regions, is also influenced by the lack of experimental data of extremely low- and high-$p_T$ regions which leads to some errors in determining the trend of $p_T$ distribution.

\begin{figure}[H]
\hskip-0.0cm {\centering
\includegraphics[width=16.0cm]{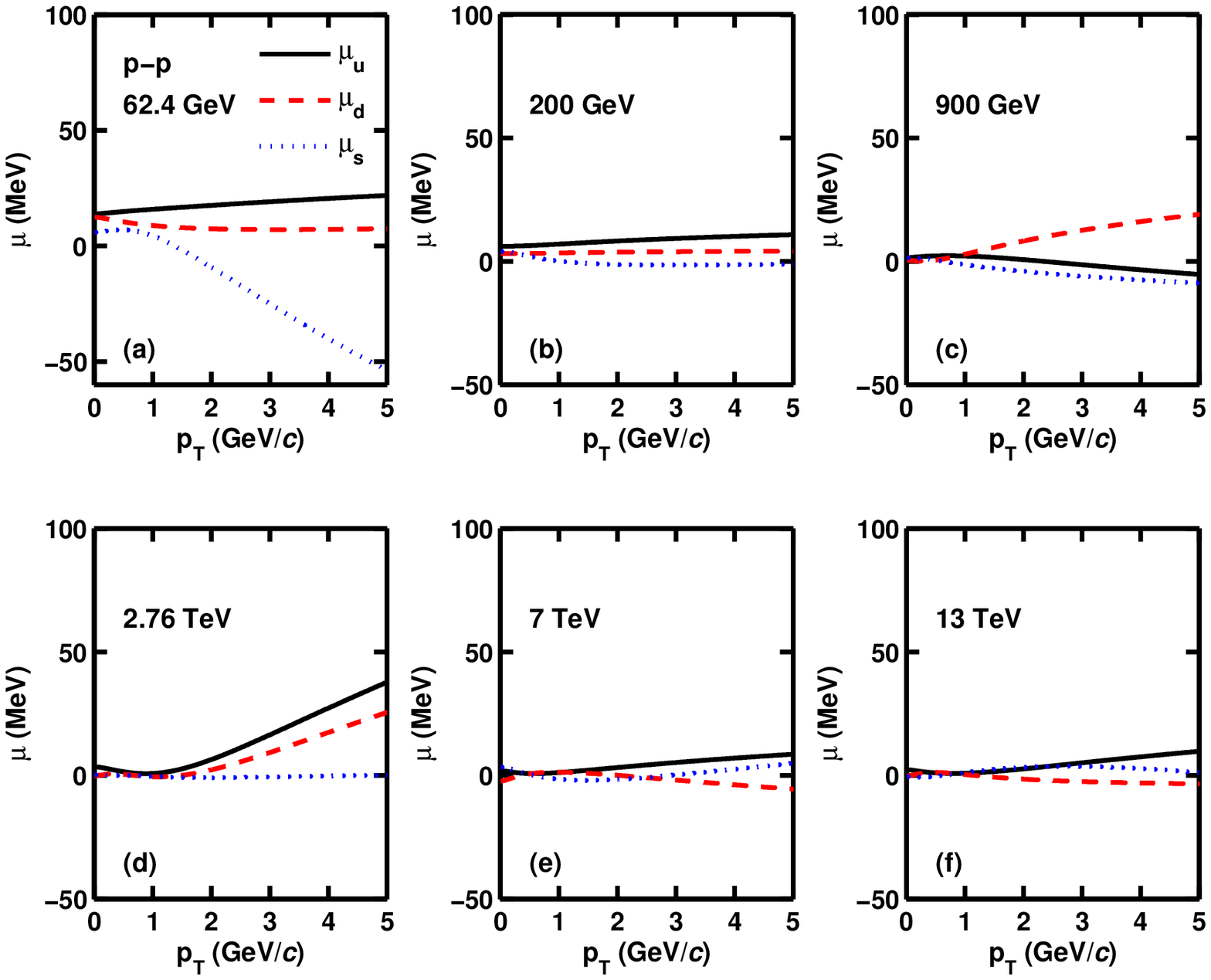}}
\vskip-0.18cm  Figure 4. Transverse momentum dependent chemical potentials, $\mu_{u}$, $\mu_{d}$, and $\mu_{s}$ of $u$, $d$, and $s$ derived from $pp$ collisions at different energies shown in panels (a)--(f). The curves are our results calculated by using Eq. (7).
\end{figure}

From that the six types of curves converge near zero in low-$p_T$ region, especially when the energy is above 900 GeV, and diverge in high-$p_T$ region, we can speculate that the yield gap between positive and negative particles is small in soft excitation process (low-$p_T$ region) and relatively large in hard scattering process (high-$p_T$ region). All the values of $\sqrt{s}$-dependent chemical potentials in Fig. 5, are nearly less than the mean values of corresponding type of chemical potentials in different transverse momenta at corresponding energy in Figs. 3 and 4. This feature indicates that the particle yield from soft excitation is greater than that from hard scattering, which means soft excitation is the main process of producing particles in $pp$ collision system.

At the same time, $\sqrt{s}$-dependent chemical potentials of light hadrons gradually slightly decrease with increase of energy from RHIC to LHC, which indicates that the density of baryon number slightly decrease with increase of energy. At different energies, $\mu_{u}$ is slightly larger than $\mu_{d}$, and $\mu_{d}$ is larger than $\mu_{s}$, which results from the differences of quark masses. With the increase of energy, $\mu_{u}$, $\mu_{d}$, and $\mu_{s}$ slightly decrease, which renders that the mean-free-path of quarks slightly increases and viscous effect slightly weakens with increase of energy. In the energy range involved, the six types of $\sqrt{s}$-dependent chemical potentials in Fig. 5 are small, especially when $\sqrt{s}$ increases to about 900 GeV, all types of chemical potentials approach to zero. Combining with Figs. 3 and 4, the range of $p_{T}$ where $\mu$ being close to zero, increases with increase of $\sqrt{s}$, which indicates that the partonic interactions gradually become larger, while hadronic degrees of freedom are wearing off with increase of $\sqrt{s}$ over a range from the RHIC to LHC.

Before summary and conclusion, we would like to point out that, in the mentioned energy range, the chemical potentials in $pp$ collisions studied in the preset work is similar to those in heavy ion collisions studied in our recent works [ 10, 11, 31], though different fitting functions or treating methods are used. The present work is more systemical on the study of the energy dependent chemical potentials in $pp$ collisions. There is no obvious inconformity in $pp$ and heavy ion collisions in terms of chemical potential. This suggests universality in particle production, as it is obtained in previous and recent studies of Sarkisyan {\it et al}. for multiplicity [46--50], but now for chemical potential as well.
\\
\begin{figure}[H]
\hskip-0.0cm {\centering
\includegraphics[width=16.0cm]{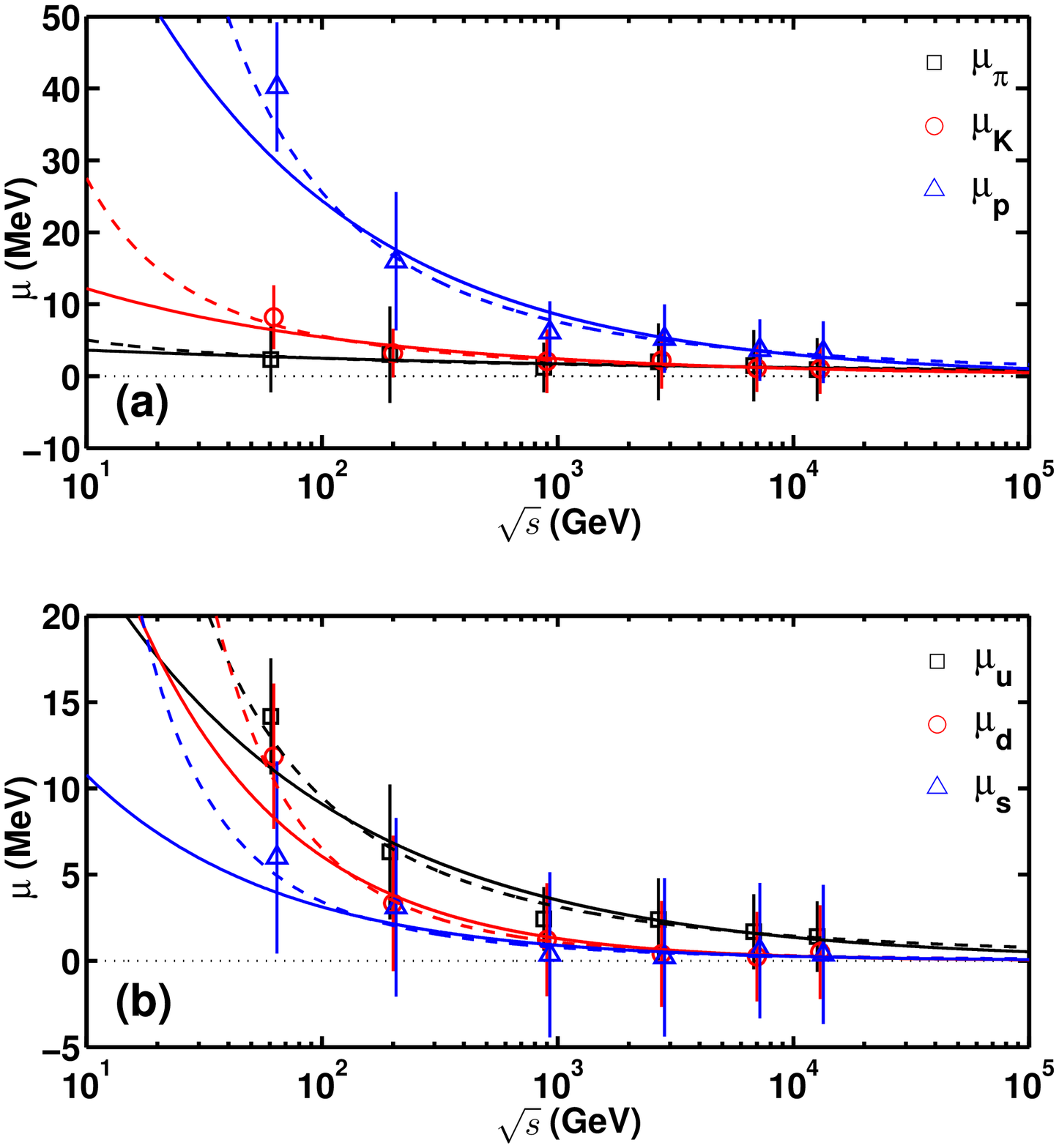}}
\vskip-0.18cm  Figure 5. $\sqrt{s}$-dependent chemical potentials of (a) light hadrons $\pi$, $K$, and $p$, and (b) light quarks $u$, $d$, and $s$. The symbols denote the calculated results according to $N_0$ and Eqs. (6) and (7), where the squares and triangles are shifted by a small amount for clarity. The solid and dashed curves are the empirical fits according to Eqs. (8) and (9), respectively.
\end{figure}

{\section{Summary and Conclusion}}

The transverse momentum spectra of final-state light flavour particles ($\pi^{\pm}$, $K^{\pm}$, $p$, and $\bar p$), produced in $pp$ collisions over an energy range from 62.4 GeV to 13 TeV, are described by a Tsallis-Pareto-type function. The experimental data recorded by the PHENIX and CMS Collaborations are well fitted by the function.

The $p_T$-dependent and $\sqrt{s}$-dependent chemical potentials of light hadrons, $\mu_{\pi}$, $\mu_{K}$, and $\mu_{p}$, and quarks, $\mu_{u}$, $\mu_{d}$, and $\mu_{s}$, are extracted from the yield ratios of negative to positive particles based on the fitting results of transverse momentum spectra and the normalization constants. The six types of $p_T$-dependent chemical potentials are close to zero in low-$p_T$ region, especially at $\sqrt{s}>900$ GeV, and are away from zero in high-$p_T$ region. Meanwhile, with increase of $\sqrt{s}$, the $\sqrt{s}$-dependent $\mu_{K}$, $\mu_{p}$, $\mu_{u}$, $\mu_{d}$, $\mu_{s}$, and $\mu_{\pi}$ seem to decrease slightly. The limiting values of the six types of chemical potentials are zero in $pp$ collisions at very high energies.

With the increase of $\sqrt{s}$ over a range from 62.4 GeV to 13 TeV, the range of $p_{T}$ corresponding to small (near zero) $p_T$-dependent chemical potentials increases, and the
$\sqrt{s}$-dependent chemical potentials slightly decrease and gradually approach zero. These features show that the higher the energy is, the greater the role of partonic interactions plays. Especially, partonic interactions possibly play a dominant role at the LHC.\\

{\bf Data Availability}

All data are quoted from the mentioned references. As a phenomenological work, this paper does not report new data.\\

{\bf Conflicts of Interest}

The authors declare that there are no conflicts of interest regarding the publication of this paper.\\

{\bf Acknowledgments}

This work was supported by the National Natural Science Foundation of China under Grant Nos. 11575103 and 11747319, the Shanxi Provincial Natural Science Foundation under Grant No.
201701D121005, and the Fund for Shanxi ``1331 Project" Key Subjects Construction.

\end{document}